# Computational assessment of incremental pyriform aperture enlargement in plastic surgery: effect on inhaled nasal airflow

**Prosun Roy, Ph.D.[1,*]; Mohammad Yeasin, M.S.[2]; Siena Shah, B.S.[3]; Elliott Bryner. Ph.D.[1]; Anil Shah, M.D.[4,5]; Saikat Basu, Ph.D.[2,*]**

[1]Department of Mechanical Engineering, Embry-Riddle Aeronautical University, 3700 Willow Creek Road, Prescott, AZ 86301, USA

[2] Department of Mechanical Engineering, South Dakota State University, Brookings, SD 57007, USA

[3]Department of Physics, Cornell University, Ithaca, New York 14850, USA

[4]Section of Otolaryngology-Head and Neck Surgery, Department of Surgery, The University of Chicago, Chicago, Illinois, USA

[5]Shah Aesthetics, 200 W Superior St, Chicago, Illinois 60654, USA.

**Corresponding authors*:

[1]Prosun Roy; Email: royp2@erau.edu; Phone: +1-928-777-3848

[2]Saikat Basu; Email: Saikat.Basu@sdstate.edu; Phone: +1-605-688-6868; ORCID iD 0000-0003-1464-8425

# Abstract

**Background.** The pyriform aperture is a critical anterior bony constriction governing nasal airflow and is clinically relevant to both functional and aesthetic nasal surgery, yet the degree of pyriform aperture enlargement (PAE) needed to capture its aerodynamic benefit is undefined. This study employs high-fidelity computational fluid dynamics (CFD) to quantify the aerodynamic effects of progressive PAE and identify a threshold beyond which further enlargement yields minimal benefit.

**Methods.** A CT-derived three-dimensional nasal airway model was digitally modified to produce baseline, 5%, 10%, 15%, and 20% increases in pyriform aperture cross-sectional area. Pressure-driven large-eddy simulations (LES) were performed at a fixed transnasal pressure difference of 11.25 Pa to quantify changes in inspiratory flow rate, total airway resistance, and local airflow characteristics.

**Results** Progressive PAE monotonically increased inspiratory flow from 15.063 to 15.252 L/min (approximately 1.25%) and reduced airway resistance from 0.6138 to 0.6028 Pa·min/L (approximately 1.80%). The aerodynamic response was strongly nonlinear, with the greatest improvement occurring between 10% and 15% enlargement. Increasing the aperture from 15% to 20% produced negligible additional improvement.

**Conclusions.** PAE improves nasal airflow by reducing the resistance imposed by the anterior bony constriction. In the anatomical model investigated, approximately 15% area enlargement represented an efficient aerodynamic threshold, after which the dominant resistance shifted toward downstream nasal structures and further enlargement produced little additional benefit. These findings support a conservative, anatomy-dependent approach to PAE and warrant validation across a broader range of patient-specific geometries.

**Level of Evidence:** No Level Assigned.

**Key Points**

- Graded virtual pyriform aperture enlargement (PAE) (5–20% of cross-sectional area) monotonically increased inspiratory flow and decreased nasal airway resistance at a constant transnasal pressure.
- The benefit was strongly nonlinear: the 10–15% step delivered approximately 64% of the total resistance reduction, whereas enlargement beyond 15% added less than 1%.
- At saturation, the dominant resistance shifted from the bony inlet to the downstream septal and turbinate region, placing the flow-limiting segment beyond the reach of further enlargement.

## Introduction

### *Anatomy and Physiology of the Nasal Valve*

The nasal airway warms, humidifies, and filters inspired air before it reaches the lower respiratory tract [1, 2]. Within this pathway, the primary regulator of the airway is the nasal valve region. As a result, the nasal valve region accounts for the majority of nasal respiratory resistance [3, 4]. The internal nasal valve is bounded by the caudal edge of the upper lateral cartilage, the septum, the nasal floor, and the head of the inferior turbinate. The external nasal valve lies near the alar rim and is formed by the alae, the crura of the lower lateral cartilages, and the columella [3, 4]. Because of its narrow cross-section, the valve region remains a key factor in nasal function.

The nasal valve is not a fixed opening. Its cross-sectional area changes with cartilage support, alar muscle activity, and swelling of the turbinate and septal tissue [5]. As air speeds up through the valve, the pressure inside drops and pulls the side walls inward. If these walls are weak, they can collapse during inspiration [6].

### *The Pyriform Aperture as a Flow-Limiting Structure*

The pyriform aperture is bounded by the maxilla laterally and inferiorly and by the nasal bones superiorly (Fig. 1a). It is the narrowest bony part of the nasal airway, so even small changes in its size can noticeably change nasal resistance [7-9]. It also underlies the alar base and midfacial contour, and is itself a target of augmentation in facial rejuvenation [10].

The classical model considers the nasal valve as a Starling resistor, in which an increase in inspiratory effort increases the negative pressure at the choke point until a critical value is reached and the elastic resistance of the wall is overcome and flow limitation occurs [11, 12]. The pyriform aperture is bony and non-collapsible and represents the fixed (static) part of this system, whereas the compliant cartilaginous valve is the dynamic part [4]. The importance of the bony inlet is clear in congenital nasal pyriform aperture stenosis (CNPAS). In this condition, a narrow aperture can cause life-threatening airway obstruction in newborns, who are primarily nasal breathers [13-15]. In a computational case study, total nasal resistance in a neonate with CNPAS was approximately eight times that of a normal neonatal airway [16], and virtual surgery has been used to model the effect of aperture enlargement in these patients [17]. Aperture size also varies between populations and sexes. For instance, Korean populations exhibit larger pyriform aperture widths than Caucasian populations [18], individuals of African descent have wider apertures still [19], and males have larger apertures than females [20-22].

### *Clinical Evaluation of Nasal Obstruction*

Nasal obstruction is a frequent clinical problem whose evaluation relies on both subjective patient reporting and objective physiological measurement [23]. Subjective tools like the Nasal Obstruction Symptom Evaluation (NOSE) questionnaire can provide validated metrics for quantifying patient discomfort. However, subjective perception does not always correlate with objective anatomical or aerodynamic findings [24, 25]. For example, patients with a paradoxical sensation of nasal obstruction may report feeling congested on the side of the nose that objectively has lower airflow resistance.

Objective measurements of nasal patency primarily use rhinomanometry and acoustic rhinometry. Acoustic rhinometry locates the minimal cross-sectional area, usually at the nasal valve or the head of the inferior turbinate, whereas rhinomanometry derives nasal resistance from transnasal pressure and airflow [23, 25]. The correlation between rhinomanometric data and subjective symptoms is subject to debate, often requiring pronounced differences in resistance to accurately reflect patient perception [23, 25-27].

### *Surgical Interventions for Nasal Valve Collapse and Pyriform Aperture Stenosis*

When medical treatment fails, surgery is often the next step. Valve surgery either widens the airway or strengthens the side walls [3]. Common interventions include the placement of spreader grafts between the upper lateral cartilages and the septum to widen the internal nasal valve angle, and alar batten grafts to support the lateral crura of the external valve [3, 4].

More recently, the bony inlet itself has become a surgical target. PAE removes bone from the maxilla to widen the opening (Fig. 1b, c). It can be performed through sublabial, endonasal, or open rhinoplasty approaches, often with an ultrasonic bone aspirator, which allows precise bone removal while minimally disturbing the surrounding soft tissues [7, 8, 28, 29]. Reported outcomes have been favorable, with significant reductions in NOSE scores and improvements in rhinomanometric airflow measurements [8, 28, 29]. Fig. 2 presents representative preoperative and 6-month postoperative appearances of two patients who underwent pyriform aperture enlargement (PAE), illustrating changes in external nasal appearance from oblique, frontal, and lateral views.

### *Computational Fluid Dynamics in Nasal Airflow Analysis*

Quantifying the exact fluid dynamic changes that resulted from miniscule anatomical modifications is a difficult task. CFD using anatomically accurate three-dimensional reconstructions usually built from computed tomography (CT) scans, allows non-invasive simulation of airflow, pressure, wall shear stress, particle deposition, and heat transfer in the nasal cavity, and permits controlled "what-if" experiments that cannot be performed in vivo [30-33]. High-resolution three-dimensional models can allow researchers to evaluate the aerodynamic impact of specific structural changes. Fluid dynamics within the human nose are characterized by constantly transitioning airflows. The narrow nasal valve speeds up incoming air into a jet that shapes the flow further downstream. Nasal resistance is shared between the valve and the turbinate region, so changes in valve size strongly affect the total pressure drop [1, 9]. Virtual-surgery studies have used CFD to estimate the effect of individual surgical steps, predict postoperative changes, and support aperture enlargement in CNPAS [16, 17, 34-38].

The degree of enlargement beyond which widening the pyriform aperture stops yielding functional benefit has not, however, been defined. Because unnecessary bone removal carries structural and aesthetic risk, this study used LES in a patient-specific nasal airway to quantify the aerodynamic effect of graded PAE and to identify that threshold.

## Materials and Methods

### *Anatomical Modeling*

The upper airway geometry was reconstructed from existing, de-identified, medical-grade CT images of a 61-year-old female with disease-free airways. The coronal slice spacing was 0.348 mm, fine enough to capture the anterior geometry of the nasal valve and pyriform aperture. The retrospective computational use of these de-identified imaging data was approved by the Institutional Review Board (IRB) of the institution of one of the corresponding authors.

The DICOM images were imported into Mimics Research 18.0 (Materialise, Plymouth, MI). The nasal airspace was segmented using a radiodensity range of −1024 to −300 Hounsfield units, followed by manual editing to correct partial-volume and connectivity artifacts [39]. The model included the nasal vestibule, the main nasal cavity with the inferior and middle turbinates, the pyriform aperture, and the nasopharynx (Fig. 3a–e). The pyriform aperture was located on sagittal, coronal, and axial views. Its lateral bony boundaries were then virtually resected to create four enlargements, defined as 5%, 10%, 15%, and 20% increases in aperture cross-sectional area relative to baseline (Fig. 3f–j).

Because the aperture is non-circular, linear width does not scale linearly with area; both the achieved cross-sectional area ($mm^2$) and a representative width (mm) were therefore considered for each configuration, together with the actual percentage area change realized, rather than a nominal label alone. Representative widths were 10.63 mm at baseline and 12.26, 12.55, 12.55, and 12.69 mm for the 5%, 10%, 15%, and 20% configurations, respectively. Width was measured as the minimum separation across the middle of the aperture periphery (red lines, Fig. 3f–j); because the 10%-to-15% resection added area above and below this line rather than at it, the measured width is identical in the two configurations (Fig. 3h, i). Enlargement was defined by area rather than width on purpose: aperture dimensions vary widely with sex and ancestry [18-20], so a percentage change in area is easier to compare across patients than an absolute width in millimeters.

### *Simulation Setup*

The reconstructed geometry was exported as stereolithography (STL) files and imported into ICEM CFD 2025 R1 (ANSYS Inc., Canonsburg, PA) for spatial discretization. Following established mesh-refinement protocols for nasal airflow [40, 41], each mesh had four prism layers (each about 0.1 mm thick, height ratio 1) along the airway walls, excluding the inlet and outlet planes, to resolve near-wall velocity gradients. The core was filled with about four million unstructured tetrahedral elements generated with the robust (Octree) method (Fig. 3k, l).

The LES scheme was used to evaluate the aerodynamic impact of the geometric modifications which resolves the large, energy-containing turbulent structures explicitly and divides the turbulent motion into large- and small-scale components [42-44]. Validation studies indicate that at restful flow a laminar or well-resolved scale-resolving model reproduces measured velocity and turbulence fields, whereas Reynolds-averaged closures are less accurate in this regime; LES was therefore selected to resolve the anterior flow faithfully across all geometries [31]. A subgrid-scale kinetic-energy transport model was invoked to represent the unresolved small-scale fluctuations [45, 46]. Simulations were run in ANSYS Fluent 2025 R1 (ANSYS Inc., Canonsburg, PA) using a segregated solver, SIMPLEC pressure–velocity coupling, and second-order upwind discretization. At each time step, convergence required continuity residuals below $10^{-4}$ and velocity residuals below $10^{-5}$. Inhalation was simulated over 0.35 s with a fixed time step of 0.0002 s, taking about 1–1.5 days of computing time per case [42]. To represent warmed intranasal air along the respiratory route, the dynamic viscosity of air was set to $1.825 \times 10^{-5}$ $kg \cdot m^{-1} \cdot s^{-1}$ and its density to 1.204 $kg \cdot m^{-3}$.

The flow was pressure-driven to mimic inspiration. A constant transnasal pressure difference of 11.25 Pa was applied between the nostrils and the outlet downstream of the nasopharynx in all five cases, so that the inspiratory flow rate and the airway resistance responded only to the aperture modification. This pressure lies within the range used in resting-inspiration nasal CFD studies, in which transnasal pressure drops of 10–20 Pa produce bilateral flows near 15 L/min; the resulting baseline flow was compared against these benchmarks as a validation check [47, 48]. Airway walls were treated as rigid, smooth, and no-slip, standard simplifications justified because separation-dominated pressure losses dominate frictional losses in this geometry, while acknowledging that they omit mucosal compliance and dynamic collapse.

# Results

## *Effects on Flow Rate and Airway Resistance*

Under the fixed transnasal driving pressure (applied between the nostril openings and the base of the outlet tube downwind from the nasopharynx), progressive enlargement of the pyriform aperture produced a consistent, monotonic improvement in nasal airflow: inspiratory volumetric flow increased and total airway resistance decreased at every stage of the series, from the baseline geometry through the 20% enlargement. The flow resistance was evaluated as $R = |\Delta P|/Q$, where $\Delta P$ is the area-weighted average total pressure drop between the nostril inlet and outlet planes, obtained from the converged solution, and Q is the volumetric flow rate. Total inspiratory flow rose from 15.063 L/min in the baseline geometry to 15.252 L/min in the 20% enlarged geometry, an increase of approximately 1.25%, while the corresponding airway resistance fell from 0.6138 to 0.6028 Pa·min/L, a reduction of approximately 1.80%.

The stepwise widening of the pyriform aperture gradually decreased the local airflow velocity in the anterior nasal region (Fig. 4a–f)). As the opening widened, there was more space for the air to pass through, and it didn't have to speed up as much in this constricted region. As a result, the high-velocity area became smaller and less intense with increasing PAE. Streamlines illustrate the change in the local flow field (Fig. 4g–j). At baseline, the constricted anterior passage produced a focused high-velocity jet immediately downstream of the pyriform aperture; after 15% enlargement, the jet was less concentrated and local velocity was lower. Because total resistance fell under the fixed pressure difference, inspiratory flow nonetheless increased, and velocity at the nasopharyngeal outlet was modestly higher than at baseline (Fig. 4k–m).

The relationship between enlargement and benefit was strongly nonlinear (Fig. 5). Of the total baseline-to-20% resistance reduction, the 0–5% step contributed about 7%, the 5–10% step about 28%, and the 10–15% step roughly 64%, whereas the final 15–20% step contributed less than 1%. The marginal benefit per increment of enlargement therefore peaked in the 10–15% step and then fell essentially to zero.

### *Identification of the Aerodynamic Saturation Point*

The nonlinear dose–response localized a distinct saturation point near 15% area enlargement. Between the 15% and 20% geometries, the aerodynamic metrics were nearly indistinguishable: volumetric flow increased by less than 0.01 L/min (from 15.242 to 15.252 L/min) and resistance decreased by less than 0.0001 Pa·min/L, accounting for under 1% of the total benefit realized across the series. Thus, the 15% enlargement accounted for almost all the aerodynamic improvement, the response being practically saturated above that. The optimum range of enlargements was 10% to 15%, the single largest gain, and further enlargement from 15% to 20% gave little additional relief. Thus, for this anatomical model, 15% area enlargement is an efficient aerodynamic threshold, the least modification to capture essentially the full benefit available from widening the bony inlet, with no meaningful functional return from further bone removal. By the saturation point, the dominant resistance had shifted from the widened bony inlet to the downstream nasal structures, so further enlargement removed bone without changing the segment that governed flow.

## Discussion

### *Aerodynamic Shifts in Nasal Resistance*

The saturation near 15% enlargement reflects a change in the location of the flow-limiting segment. At baseline, the narrow pyriform aperture acts as a bottleneck, and relieving it improves flow until the aperture is no longer the narrowest constriction. Total nasal resistance is a composite of losses across the valve region and the main nasal cavity, including the turbinates, whose highly vascular mucosa can engorge substantially [5, 9]. Once the bony inlet is wide enough, the turbinate region becomes rate-limiting, and further widening of the aperture cannot meaningfully reduce total resistance.

### *Clinical Implications for Surgical Planning*

The findings from this study can carry significant implications for facial plastic and reconstructive surgeries requiring PAE. The PAE improves both NOSE scores and rhinomanometric airflow [8, 28], but removing bone from the maxilla is not without risk. Aggressive resection can compromise support of the upper lateral cartilages and lead to secondary collapse of the middle nasal vault, and excessive lateral or inferior drilling can injure the anterior superior alveolar nerve, causing dental numbness or pain [20]. Because essentially the entire aerodynamic benefit was captured at the saturation point, these results argue for a conservative, tissue-preserving target: enlarge the aperture until it is no longer flow-limiting, and no further. In aesthetic and functional rhinoplasty, where alar base position and midfacial contour matter, this confines bone removal to what produces a functional gain. In this model the saturation point lay near 15% area enlargement, but the transferable principle is the criterion itself rather than a fixed percentage, because the bone volume needed to reach it will vary with each patient's baseline anatomy.

The results also help when PAE does not fully relieve obstruction. Once the inlet is wide enough, most of the remaining resistance lies further inside the nose, so persistent obstruction is more likely due to the septum or turbinates. Adding septoplasty or submucosal turbinate reduction may then help, consistent with virtual-surgery studies showing that these procedures provide most of the resistance reduction in multilevel nasal surgery. Resistance alone, however, is a limited basis for judging surgical outcomes. Patients do not sense resistance directly; they sense airflow through cooling of the nasal mucosa, detected by cold-sensitive trigeminal nerve endings. This is why menthol can make breathing feel easier without changing nasal shape, and why CFD-based heat-transfer measures often match patient symptoms better than resistance [11, 49-51].

A 1.8% reduction in resistance therefore need not translate directly into perceived improvement. Conjugate heat-transfer simulations quantifying mucosal heat flux and cooled surface area across aperture sizes would show whether perceived patency plateaus at the same point as mechanical resistance.

### *Complexities of Unsteady Flow and Morphological Variations*

This study applied a constant pressure difference, which isolates the effect of geometry. Real breathing, however, involves changing flow rates, and during these changes turbulence and wall shear stress can differ from steady-flow predictions[52]. Simulations of full breathing cycles would refine these results. The study also used a single anatomical model. Because aperture size and shape vary with sex and ancestry, the baseline resistance and the amount of widening reported here are specific to this subject, although the overall trend is expected to apply more widely. Congenitally narrow, post-traumatic, and normal adult apertures start from different conditions, so the bone removal needed to reach saturation will differ between patients. Defining the endpoint as a percentage of area, and by the point where the main resistance moves downstream, makes the criterion easier to apply across patients, but it needs to be tested in multiple airway models. The rigid-wall assumption also ignores dynamic collapse of the cartilaginous valve, which aperture enlargement may affect. Finally, widening the inlet changes the airflow patterns that control where inhaled particles deposit, so these changes may also affect intranasal drug delivery [53-56], and, more broadly, the regional deposition of any inhaled material, including pathogenic particulates [57].

## Conclusion

Pressure-driven LES in a patient-specific nasal airway showed that PAE monotonically decreases airway resistance and increases inspiratory flow. The benefit was concentrated between 10% and 15% area enlargement; beyond 15%, enlargement added negligible benefit because the dominant resistance shifted to downstream nasal structures. In this model, approximately 15% enlargement was an efficient aerodynamic threshold, providing an objective, anatomy-dependent target that preserves the functional gain while limiting the bone resection that risks cartilage destabilization, nerve injury, and aesthetic change. Validation across multiple patient-specific geometries is warranted.

## Acknowledgements and Disclaimer

The authors used artificial intelligence–assisted language tools, including Claude and ChatGPT, to refine the manuscript for grammatical accuracy, clarity, and overall readability. The use of these tools was limited to editorial and linguistic improvements and had no role in the development of the scientific methodology, analysis of the data, interpretation of the results, or formulation of the conclusions.

## Data availability statement

All essential information is contained in the article. Supplementary Material (including anatomical geometries, simulation files, postprocessing spreadsheets, and programming codes) are available via the open access repository figshare. The reader may also contact the corresponding author for any relevant data.

## Compliance with Ethical Standards

***Conflict of Interest***: One co-author is a practicing facial plastic surgeon and is associated with for-profit medical practice. That co-author did not receive any compensation for this work. The remaining authors declare no conflict of interest.

***Funding*****:** Funding information is provided on the title page to preserve blinding.

***Ethical Approval:*** The use of previously collected and anonymized human imaging data for computational research was approved by the Institutional Review Board (IRB) of the institution of one of the corresponding authors. All procedures were conducted in accordance with applicable local regulations and institutional guidelines. In accordance with national legislation and institutional requirements, written informed consent was not required from the participants, their legal guardians, or next of kin.

***Human and Animal Rights*****:** This article does not contain any studies of animals performed by any of the authors.

***Informed Consent*****:** Informed consents were obtained from patients for the publication of identifiable photographs shown in Fig. 2

## References

1. Doorly DJ, Taylor DJ, Schroter RC (2008) Mechanics of airflow in the human nasal airways. Respir Physiol Neurobiol 163: 100–110
2. Yuk J, Akash MMH, Chakraborty A, Basu S, Chamorro LP, Jung S (2023) Morphology of pig nasal structure and modulation of airflow and basic thermal conditioning. Integrative and comparative biology 63: 304–314
3. Spielmann PM, White PS, Hussain SS (2009) Surgical techniques for the treatment of nasal valve collapse: a systematic review. Laryngoscope 119: 1281–1290
4. Tasca I, Ceroni Compadretti G, Sorace F (2013) Nasal valve surgery. Acta Otorhinolaryngol Italica 33: 196
5. Cole P (2003) The four components of the nasal valve. American Journal of Rhinology 17: 107–110
6. Barrett DM, Casanueva FJ, Cook TA (2016) Management of the nasal valve. Facial Plastic Surgery Clinics 24: 219–234
7. Roy S, Iloreta AM, Bryant LM, Krein HD, Pribitkin EA, Heffelfinger RN (2015) Piriform aperture enlargement for nasal obstruction. The Laryngoscope 125: 2468–2471
8. Shah AR, Grande DP, Buckner H, Nordgren R (2024) Pyriform aperture enlargement through a rhinoplasty approach In: Aesthetic Surgery Journal Open Forum Oxford University Press US, pp ojae115
9. O'Neill G, Tolley NS (2019) The complexities of nasal airflow: theory and practice. Journal of Applied Physiology 127: 1215–1223
10. Pessa JE, Peterson ML, Thompson JW, Cohran SC, Garza JR (1999) Pyriform augmentation as an ancillary procedure in facial rejuvenation surgery. Plastic & Reconstructive Surgery 103: 683–686
11. Barbarite E, Gadkaree SK, Melchionna S, Zwicker D, Lindsay RW (2021) Computational fluid dynamics modeling of nasal obstruction and associations with patient-reported outcomes. Plastic & Reconstructive Surgery 148: 592e–600e
12. Le TB, Moghaddam MG, Woodson BT, Garcia GJ (2019) Airflow limitation in a collapsible model of the human pharynx: physical mechanisms studied with fluid-structure interaction simulations and experiments. Physiological reports 7: e14099
13. Belden CJ, Mancuso AA, Schmalfuss IM (1999) CT features of congenital nasal piriform aperture stenosis: initial experience. Radiology 213: 495–501

14. Moreddu E, Le Treut-Gay C, Triglia J, Nicollas R (2016) Congenital nasal pyriform aperture stenosis: elaboration of a management algorithm from 25 years of experience. International journal of pediatric otorhinolaryngology 83: 7–11
15. Wormald R, Hinton-Bayre A, Bumbak P, Vijayasekaran S (2015) Congenital nasal pyriform aperture stenosis 5.7 mm or less is associated with surgical intervention: a pooled case series. International Journal of Pediatric Otorhinolaryngology 79: 1802–1805
16. Patel TR, Li C, Krebs J, Zhao K, Malhotra P (2018) Modeling congenital nasal pyriform aperture stenosis using computational fluid dynamics. International journal of pediatric otorhinolaryngology 109: 180–184
17. Moreddu E, Meister L, Médale M, Nicollas R (2024) Computational fluid dynamics (CFD), virtual rhinomanometry, and virtual surgery for neonatal congenital nasal pyriform aperture stenosis. International Journal of Pediatric Otorhinolaryngology 182: 112025
18. Hwang T-S, Song J, Yoon H, Cho B-P, Kang H-S (2005) Morphometry of the nasal bones and piriform apertures in Koreans. Annals of Anatomy-Anatomischer Anzeiger 187: 411–414
19. Ofodile FA (1994) Nasal bones and pyriform apertures in blacks. Annals of plastic surgery 32: 21–26
20. Bedre Duygu O, Govsa F, Gokmen G (2026) Morphometric and morphological assessment of the piriform aperture in dry crania: anatomical observations and surgical implications. Surgical and Radiologic Anatomy 48: 149
21. Sarač-Hadžihalilović A, Ajanović Z, Hasanbegović I, Šljuka S, Rakanović-Todić M, Aganović I, Prazina I, Kapo SM, Hadžiselimović R (2022) Analysis of gender differences on pyriform aperture of human skulls using geometric morphometric method. Folia Morphologica 81: 707–714
22. Kabakci AG, Sema Polat, Mahmut Öksüzler, Fatma Yasemin Öksüzler, Ahmet Hilmi Yücel, AG KABAKCI, S. POLAT, M. ÖKSÜZLER, FY ÖKSÜZLER, and AH YÜCEL (2020) Determination of the piriform aperture morphometry and golden ratio in healthy Turkish subjects: a CT study. International Journal of Morphology 38: 444–447
23. André R, Vuyk H, Ahmed A, Graamans K, Nolst Trenité G (2009) Correlation between subjective and objective evaluation of the nasal airway. A systematic review of the highest level of evidence. Clinical otolaryngology 34: 518–525
24. Stewart MG, David L. Witsell, Timothy L. Smith, Edward M. Weaver, Bevan Yueh, and Maureen T. Hannley (2004) Development and validation of the Nasal Obstruction Symptom Evaluation (NOSE) scale. Otolaryngol Head Neck Surgery 130: 157–163
25. Thulesius H, Cervin A, Jessen M (2012) The importance of side difference in nasal obstruction and rhinomanometry: a retrospective correlation of symptoms and rhinomanometry in 1000 patients. Clinical Otolaryngology 37: 17–22
26. Schumacher MJ (2004) Nasal dyspnea: the place of rhinomanometry in its objective assessment. American Journal of Rhinology 18: 41–46
27. Gaberino C, Rhee JS, Garcia GJ (2017) Estimates of nasal airflow at the nasal cycle mid-point improve the correlation between objective and subjective measures of nasal patency. Respiratory physiology & neurobiology 238: 23–32
28. Sofia OB, Castro NP, Katsutani FS, Mitre EI, Dolci JE (2018) Evaluation of pre-and post-pyriform plasty nasal airflow. Brazilian journal of otorhinolaryngology 84: 351–359
29. Hajem H, Botter C, Al Omani M, Sounthakith V, de Bressieux E, Benkhatar H (2021) Pyriform aperture enlargement for internal nasal valve obstruction in adults: systematic review and surgical classification. Otolaryngology–Head and Neck Surgery 165: 745–750
30. Lee HP, Gordon BR (2012) Impacts of fluid dynamics simulation in study of nasal airflow physiology and pathophysiology in realistic human three-dimensional nose models. Clinical and experimental otorhinolaryngology 5: 181–187

31. Li C, Jianbo Jiang, Haibo Dong, and Kai Zhao (2017) Computational modeling and validation of human nasal airflow under various breathing conditions. Journal of Biomechanics 64: 59–68
32. Yeasin M, Akash MMH, Malakar A, Borojeni AA, Tummala A, Wu J, Bennett WD, Bodnar WM, Kimbell JS, Chakravarty A, Port JR, Basu S (2025) Decoding the mechanophysiology for inhaled onset of smallpox with model-based implications for mpox spread. Scientific Reports 16: 2142
33. Basu S (2025) On the mechanics of inhaled bronchial transmission of pathogenic microdroplets generated from the upper respiratory tract, with implications for downwind infection onset. Plos one 20: e0335962
34. Rhee JS, Cannon DE, Frank DO, Kimbell JS (2012) Role of virtual surgery in preoperative planning: assessing the individual components of functional nasal airway surgery. Archives of facial plastic surgery 14: 354–359
35. Burgos MA, Bastir M, Pérez-Ramos A, Sanz-Prieto D, Heuzé Y, Maréchal L, Esteban-Ortega F (2024) Assessing nasal airway resistance and symmetry: An approach to global perspective through computational fluid dynamics. International Journal for Numerical Methods in Biomedical Engineering 40: e3830
36. Frank-Ito DO, Kimbell JS, Laud P, Garcia GJ, Rhee JS (2014) Predicting postsurgery nasal physiology with computational modeling: current challenges and limitations. Otolaryngology--Head and Neck Surgery 151: 751–759
37. Moghaddam MG, Garcia GJ, Frank-Ito DO, Kimbell JS, Rhee JS (2020) Virtual septoplasty: a method to predict surgical outcomes for patients with nasal airway obstruction. International journal of computer assisted radiology and surgery 15: 725–735
38. Charles D, Harrison L, Hassanipour F, Hallac RR (2024) Nasal airflow dynamics following LeFort I advancement in cleft nasal deformities: a retrospective preliminary study. Diagnostics 14: 1294
39. Basu S, Frank-Ito DO, Kimbell JS (2018) On computational fluid dynamics models for sinonasal drug transport: Relevance of nozzle subtraction and nasal vestibular dilation. International journal for numerical methods in biomedical engineering 34: e2946
40. Frank-Ito DO, Wofford M, Schroeter JD, Kimbell JS (2016) Influence of mesh density on airflow and particle deposition in sinonasal airway modeling. Journal of aerosol medicine and pulmonary drug delivery 29: 46–56
41. Basu S, Witten N, Kimbell JS (2017) Influence of localized mesh refinement on numerical simulations of post-surgical sinonasal airflow. J Aerosol Med Pulm Drug Deliv 30: A14
42. Ghahramani E, Abouali O, Emdad H, Ahmadi G (2017) Numerical investigation of turbulent airflow and microparticle deposition in a realistic model of human upper airway using LES. Computers & Fluids 157: 43–54
43. Huang X, Francis I, Saha G, Rahman MM, Saha SC (2024) Large eddy simulation-based modeling of cold-air inhalation from nasal cavities to the distal lung: Insights for athlete health and performance. Results in Engineering 23: 102475
44. Longest PW, Vinchurkar S (2007) Validating CFD predictions of respiratory aerosol deposition: effects of upstream transition and turbulence. Journal of biomechanics 40: 305–316
45. Baghernezhad N, Abouali O (2010) Different SGS models in large eddy simulation of 90 square cross-section bends. Journal of Turbulence: N50
46. Farnoud A, Tofighian H, Baumann I, Garcia GJ, Schmid O, Gutheil E, Rashidi MM (2020) Large eddy simulations of airflow and particle deposition in pulsating bi-directional nasal drug delivery. Physics of Fluids 32
47. Van Strien J, Shrestha K, Gabriel S, Lappas P, Fletcher DF, Singh N, Inthavong K (2021) Pressure distribution and flow dynamics in a nasal airway using a scale resolving simulation. Physics of Fluids 33

48. Borojeni AA, Garcia GJ, Moghaddam MG, Frank-Ito DO, Kimbell JS, Laud PW, Koenig LJ, Rhee JS (2020) Normative ranges of nasal airflow variables in healthy adults. International journal of computer assisted radiology and surgery 15: 87–98
49. Zhao K, Blacker K, Luo Y, Bryant B, Jiang J (2011) Perceiving nasal patency through mucosal cooling rather than air temperature or nasal resistance. PLoS One 6: e24618
50. Bailey RS, Casey KP, Pawar SS, Garcia GJ (2017) Correlation of nasal mucosal temperature with subjective nasal patency in healthy individuals. JAMA facial plastic surgery 19: 46–52
51. Na Y, Kim Y-J, Kim HY, Jung YG (2022) Improvements in airflow characteristics and effect on the NOSE score after septoturbinoplasty: a computational fluid dynamics analysis. PLoS One 17: e0277712
52. Ariyaratne C, He S, Vardy AE (2010) Wall friction and turbulence dynamics in decelerating pipe flows. Journal of Hydraulic Research 48: 810–821
53. Basu S, Holbrook LT, Kudlaty K, Fasanmade O, Wu J, Burke A, Langworthy BW, Farzal Z, Mamdani M, Bennett WD (2020) Numerical evaluation of spray position for improved nasal drug delivery. Scientific reports 10: 10568
54. Farnoud A, Tofighian H, Baumann I, Martin AR, Rashidi MM, Menden MP, Schmid O (2021) Pulsatile bi-directional aerosol flow affects aerosol delivery to the intranasal olfactory region: a patient-specific computational study. Frontiers in Pharmacology 12: 746420
55. Liu Y, Wu D (2023) Bi-directional nasal drug delivery systems: A scoping review of nasal particle deposition patterns and clinical application. Laryngoscope Investigative Otolaryngology 8: 1484–1499
56. Hossain MT, Malakar A, Yeasin M, O'Connell W, Akash MMH, Borojeni AA, Samanta D, Williams G, Reineke J, Farias G, Jung S, Suman J, Basu S (2025) Mechanics-guided parametric modeling of intranasal spray devices and formulations for targeted drug delivery to the nasopharynx. Frontiers in Drug Delivery 5: 1721960
57. Romanos Martinez P, Rohde M, Yeasin M, Basu S, Sieben C, Port JR (2026) From viral infection to emission: host barriers shaping airborne transmission. Journal of Virology: e00753–00726

## Figures (see next)

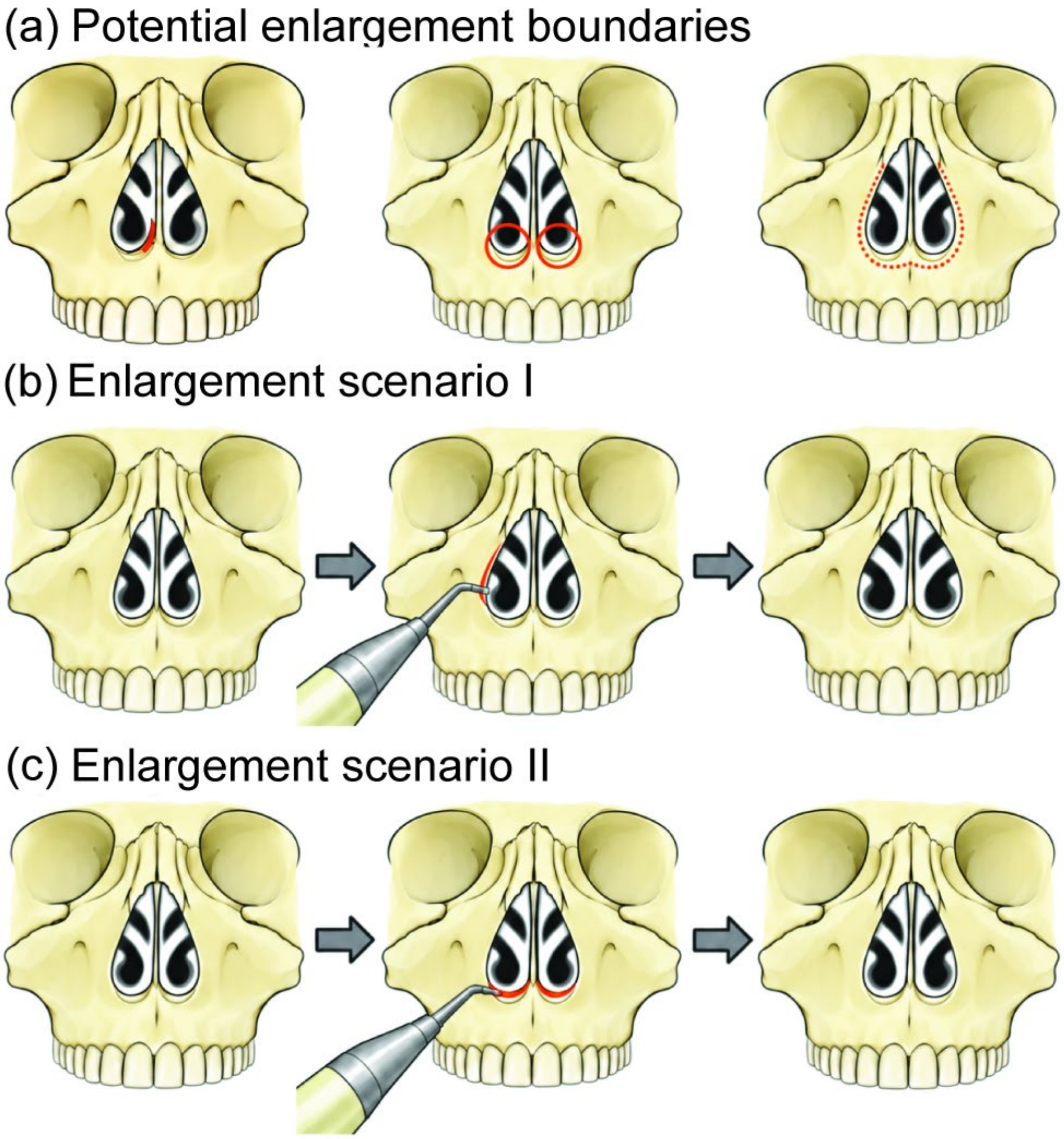

**Fig. 1:** Schematic illustration of the anatomical concept and potential surgical approaches for pyriform aperture enlargement (PAE). (a) Frontal view of the bony pyriform aperture illustrating three potential enlargement boundaries: the base of the nasal septum, the inferior bony margin, and the overall aperture perimeter, indicated by red-orange markings. (b) Enlargement scenario I, showing baseline anatomy, targeted bone removal along the lateral margin, and the resulting lateral widening of the aperture. (c) Enlargement scenario II, illustrating baseline anatomy, targeted bone removal along the inferior margin (nasal floor), and the resulting inferior enlargement of the aperture. Arrows indicate the sequence of each surgical scenario. The illustration was recreated and graphically refined with the assistance of OpenAI's ChatGPT image-generation tool, based on an author-provided schematic.

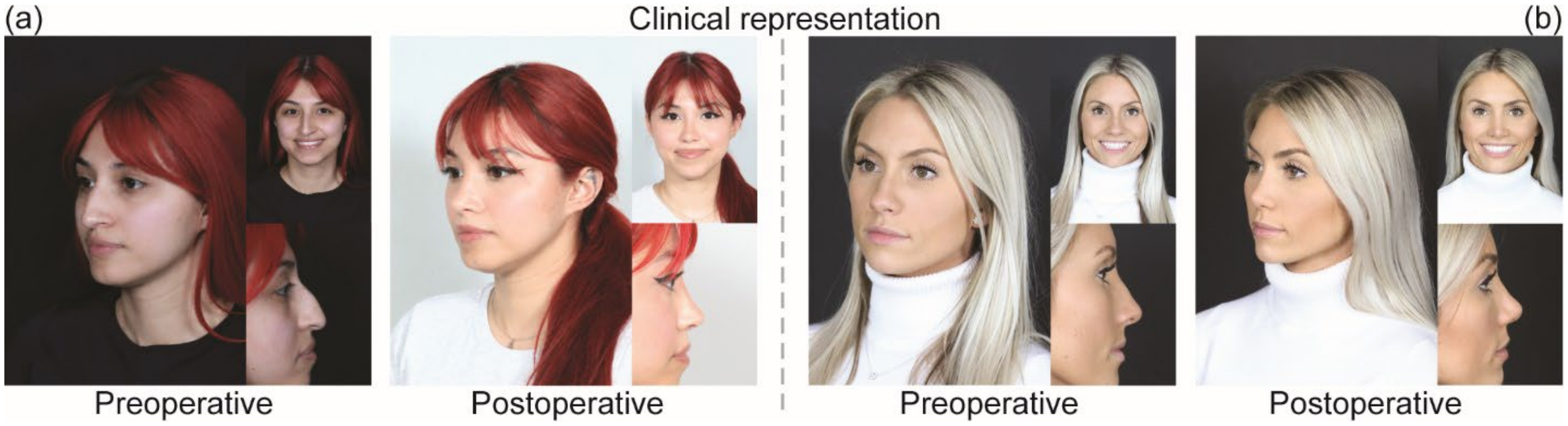

**Fig. 2.** Representative clinical photographs illustrating preoperative and postoperative external nasal appearances following pyriform aperture enlargement (PAE) in two patients. (a–b) Preoperative and 6-month postoperative appearances of two representative patients, shown in oblique (main images), frontal (upper insets), and lateral (lower insets) views, highlighting changes in external nasal contour following surgery. Clinical photographs courtesy of a co-author's clinical practice; reproduced with permission.

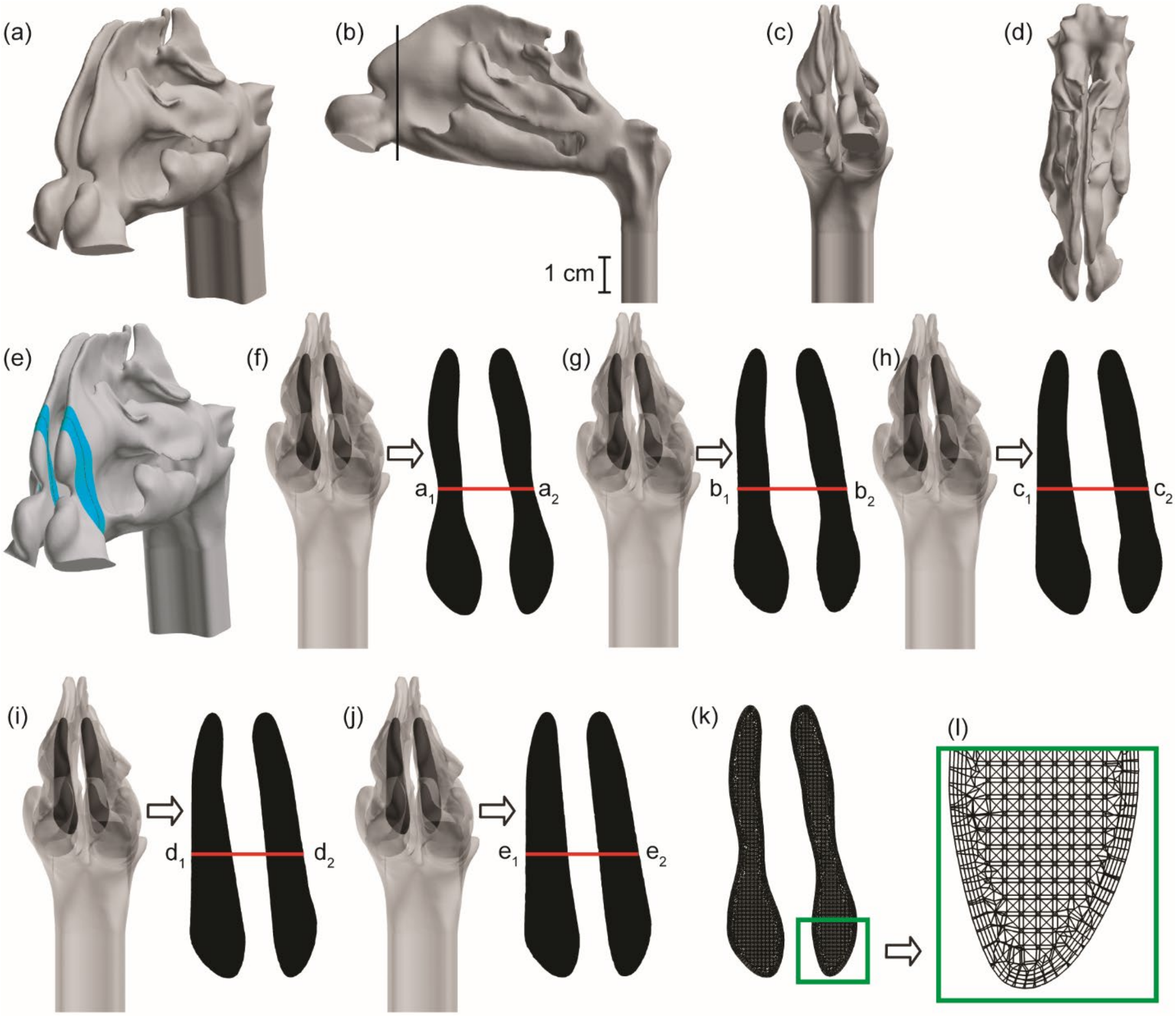

**Fig. 3**: Anatomical upper airway reconstruction and systematic enlargement of the pyriform aperture. (a–d) Three-dimensional computational model of the human nasal airway shown from different anatomical orientations; the vertical line in (b) marks the coronal plane through the pyriform aperture used for the cross-sections in (f–j). (e) Position of the pyriform aperture highlighted in the anterior nasal airway. (f–j) Sequential geometric alteration of the pyriform aperture depicting the baseline configuration and 5%, 10%, 15%, and 20% increases in cross-sectional area, respectively. For each configuration, the airway model is shown alongside the extracted cross-sectional silhouette at the aperture plane (black), on which the red line ($a_1a_2$–$e_1e_2$) marks the representative width, defined as the minimum separation measured across the middle of the aperture periphery. (k) Computational mesh shown at the nasal airway cross-section through the pyriform aperture, with the full airway model consisting of approximately 4.0 million tetrahedral elements. (l) Magnified view of the region indicated in (k), showing the near-wall mesh structure and four layers of prismatic cells generated along the airway walls.

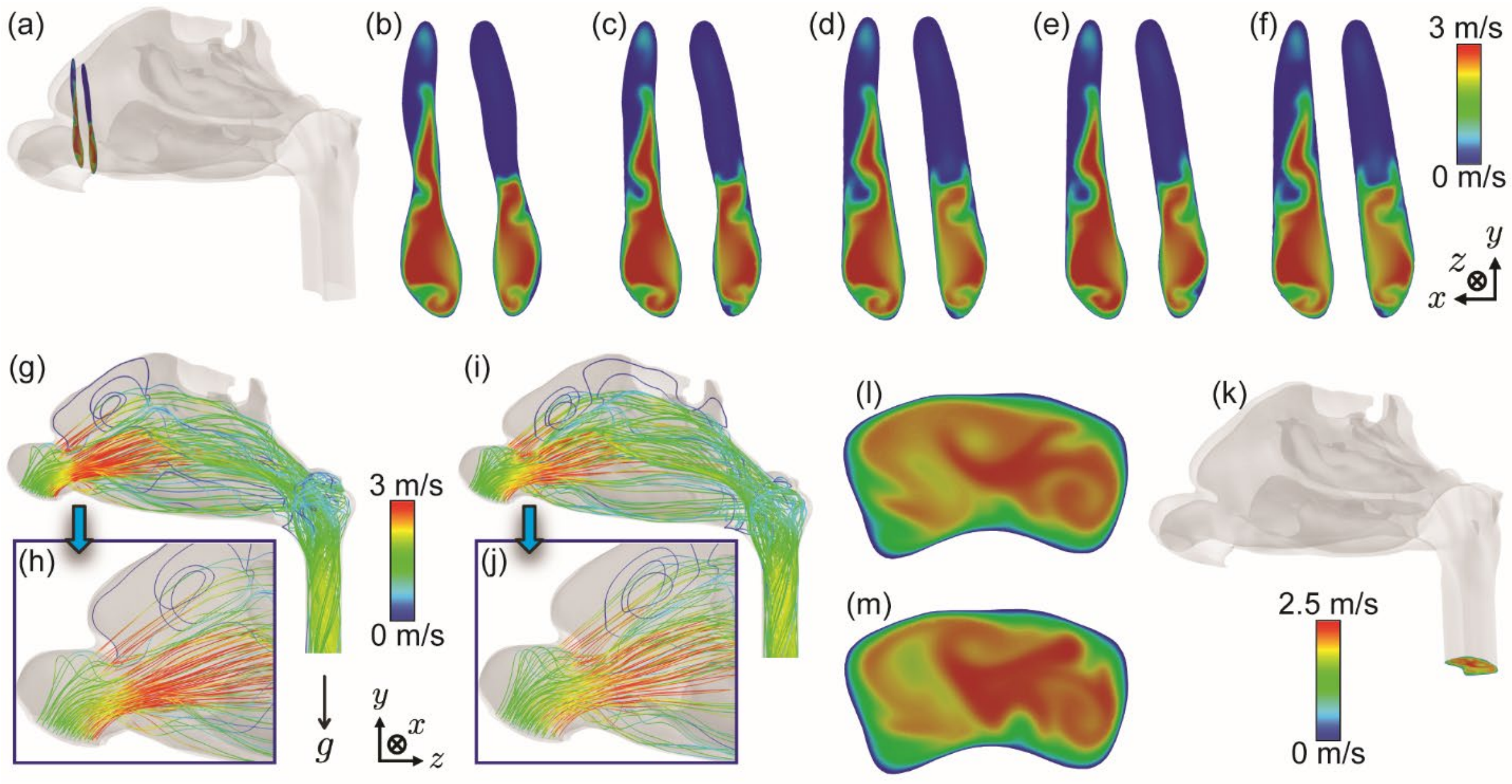

**Fig. 4**: Effect of PAE on local and global nasal airflow characteristics. (a) Nasal airway geometry showing the location of the cross-sectional plane used to evaluate velocity-magnitude contours at the pyriform aperture. (b–f) Velocity-magnitude contours at the pyriform aperture for the baseline, 5%, 10%, 15%, and 20% enlargement cases, respectively, illustrating the progressive modification of the local flow field with aperture enlargement. (g) and (i) Velocity-magnitude-colored streamlines for the baseline and 15% enlargement geometries, respectively. (h) and (j) Corresponding magnified views of the anterior nasal region, highlighting the reduction in local airflow velocity near the pyriform aperture after 15% enlargement. (k) Nasal airway geometry indicating the location of the outlet plane used for downstream velocity analysis. (l) and (m) Velocity-magnitude contours at the outlet for the baseline and 15% enlargement cases, respectively. Although local velocity near the enlarged pyriform aperture decreases

because of the increased cross-sectional area, the 15% enlargement produces a higher velocity at the outlet as a consequence of the increased overall inspiratory flow rate under the imposed constant pressure difference.

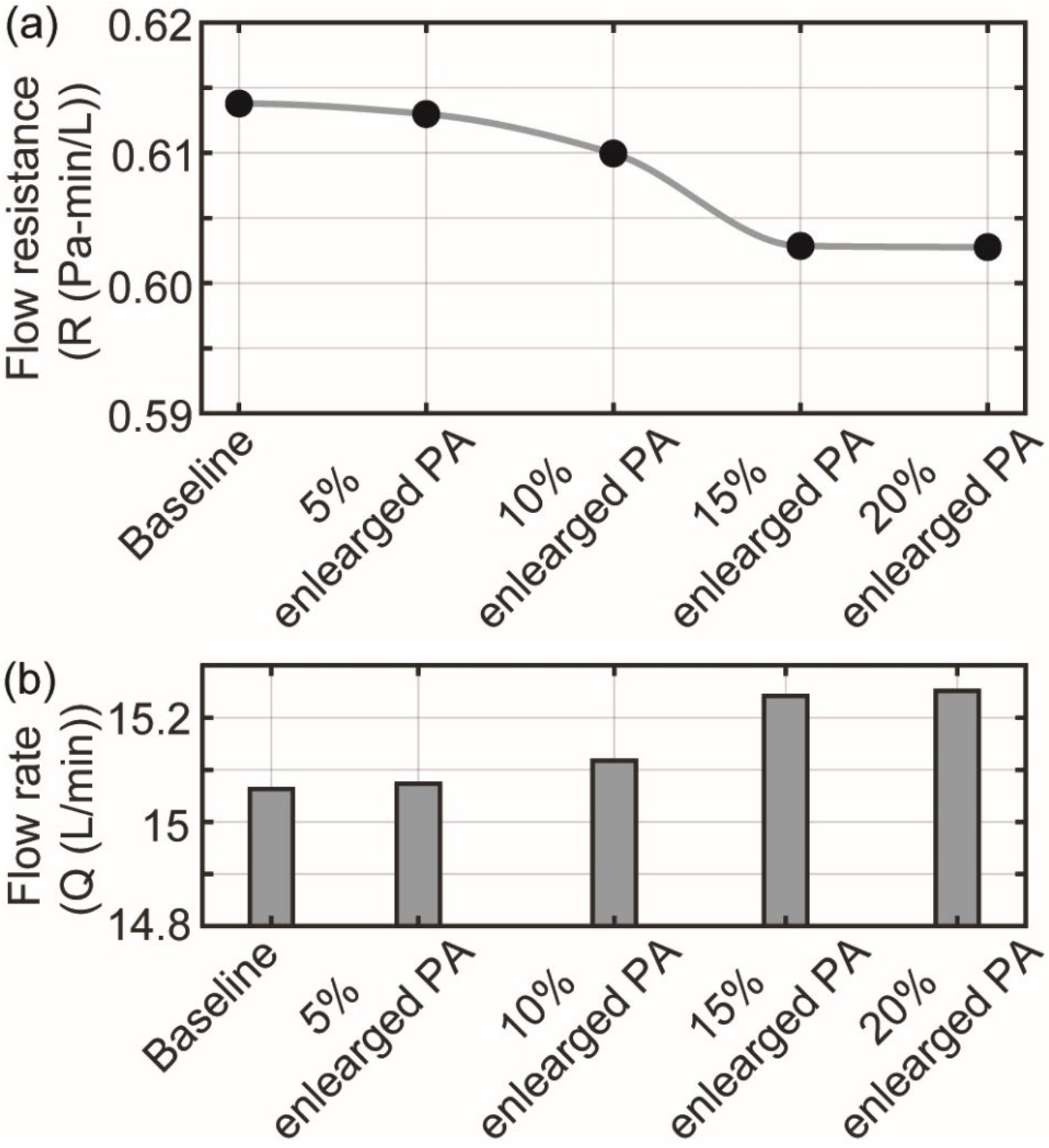

**Fig. 5**: Effect of enlargement of the pyriform aperture on nasal airflow dynamics. (a) Variation in total nasal airflow resistance and (b) inspiratory volumetric flow rate for the baseline, with 5%, 10%, 15%, and 20% PAE configurations at a constant transnasal pressure difference of 11.25 Pa. Progressive enlargement decreased airflow resistance and increased inspiratory flow, with the aerodynamic response reaching saturation at ~15% enlargement.